# NOT THE EARTH, BUT ITS ORBIT:
## ANDRÉ TACQUET AND THE QUESTION OF STAR SIZES IN A HELIOCENTRIC UNIVERSE

Christopher M. Graney

Specola Vaticana (Rome, Tucson)

Jefferson Community & Technical College (Louisville, KY 40272 USA)

christopher.graney@kctcs.edu

ABSTRACT

This paper consists of a translation of André Tacquet's discussion of the question of sizes of stars in a heliocentric universe, as published in his posthumous *Opera Mathematica* of 1668, along with introductory material and analysis. While Robert Hooke mentions Tacquet as one of the "great Anti-copernicans", who argued the question of star sizes against the heliocentric theory with "great vehemency and insulting", Tacquet's discussion has received only scant attention. The kernel of Tacquet's argument is that the absence of any detectable parallax in the stars, combined with the measured apparent sizes of the stars, means that, in a heliocentric universe, the sizes of stars compare to the size of Earth's orbit via the same proportion that they compare to the size of the Earth in a geocentric universe. The translated material presents this argument in a straightforward manner, insulting absent.

Robert Hooke in his 1674 *An Attempt to Prove the Motion of the Earth* called attention to the ideas of one of the people he identified as being among the "great anti-Copernicans": André Tacquet. Hooke was describing—

> ...a grand objection alledged by divers of the great *Anti-copernicans* with great vehemency and insulting; amongst which we may reckon *Ricciolus* and *Tacquet*, who would fain make the apparent Diameters of the Stars so big, as that the body of the Star should contain the great Orb [Earth's orbit] many times, which would indeed swell the Stars to a magnitude vastly bigger then the Sun, thereby hoping to make it seem so improbable, as to be rejected by all parties.[1]

I have discussed in some depth the anti-Copernican work of the other person Hooke identifies as a great anti-Copernican, Giovanni Battista Riccioli.[2] This paper will provide a brief introduction to Tacquet, and a translation of his star size argument against the Copernican heliocentric theory—the grand objection noted by Hooke.

Tacquet lived from 1612 to 1660. Like Riccioli, he was a priest, and in the order of the Society of Jesus. References to Tacquet are not common in recent scholarship. Perhaps this is in part because, in the words of G. H. W. Vanpaemel (in a piece on "Jesuit Science in the Spanish Netherlands"), Tacquet's life was "was utterly uneventful; he apparently never ventured outside the borders of his native province". It may also be because, according to Vanpaemel, while Tacquet produced original work, much of his effort was spent on teaching mathematics, and on producing work for teaching.[3] However, Tacquet's version of the grand anti-Copernican objection has not been wholly overlooked. Vanpaemel writes:

> Elaborating on a well-known argument against heliocentrism, Tacquet proved that in the Copernican hypothesis the proportion of the dimensions of the fixed stars to the distance earth-sun, would be equal to the proportion of the dimensions of the same stars to the radius of the earth in the geocentric hypothesis. In the Copernican hypothesis therefore,

---

[1] (Hooke 1674, 26)
[2] (Graney 2015)
[3] (Vanpaemel 2003, 406)

Graney/Tacquet, page 3 of 16

the stars needed to be much larger and heavier than in the traditional view, a conclusion which conflicted with intellectual economy.[4]

Here Vanpaemel cites Tacquet's posthumous *Opera Mathematica*.[5] However quiet and teaching-focused Tacquet may have been, Hooke was aware of his work, and thought that he stated his argument with force.

Tacquet's argument is based on the apparent sizes of stars. Astronomers from Ptolemy to Tycho Brahe had determined the apparent diameters of the more prominent (first magnitude) fixed stars to be roughly a fifteenth the apparent diameter of the moon. The wandering stars (planets) had similar apparent diameters.[6] The advent of the telescope prompted a reassessment of the apparent diameters of both fixed and wandering stars.

For example, consider the case of the wandering star Venus. A keen eye indicated the apparent diameter of Venus to be approximately one tenth the apparent diameter of the moon. But through the telescope Venus's disk showed a smaller, significantly variable apparent diameter, and phases (Figures 1, 2). The thought was that the telescope stripped away the glare or "spurious rays" from Venus, revealing the true size and appearance of that wandering star.

The telescope was thought to also do the same thing for fixed stars. The apertures used on telescopes of the seventeenth century revealed fixed stars to be distinct disks (Figure 3). Consider the following observation recorded by John Flamsteed, the first English Astronomer Royal:

> 1672, October 22. When Mercury was about 10 deg. high, I observed him in the garden with my longer tube (of 14 foot); but could not with it see the fixa [fixed star] (near him), the daylight being too strong; only I noted his diameter 45 parts = 16″ [seconds of arc], or a little less; for, turning the tube to Sirius, I found his diameter 42 parts = 15″, which I judged equal to Mercury's. The aperture on the object-glass was ¾ of an inch: so that Sirius was well deprived of spurious rays, and shined not turbulently, but as sedate as Mercury; the limbs of both well defined, but Sirius best.

---

[4] (Vanpaemel 2003, 409)
[5] (Vanpaemel 2003, 429-30)
[6] (Van Helden 1985, 27, 30, 32, 50)

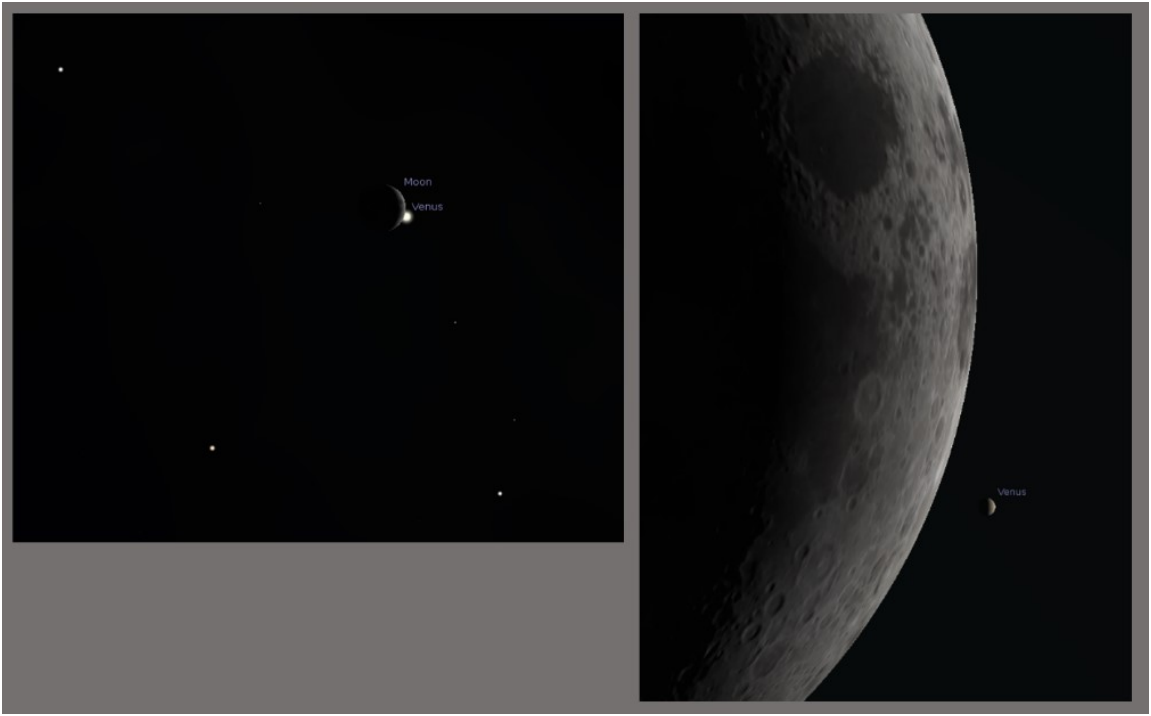

**Figure 1. The moon and Venus, seen from the Kamchatka peninsula in November of 2021, as simulated via the Stellarium planetarium app. Left is the view with the eye, right is the view with the telescope. Note that the size of Venus relative to the moon is much smaller in the right-hand image, and that Venus shows a phase like the moon.**

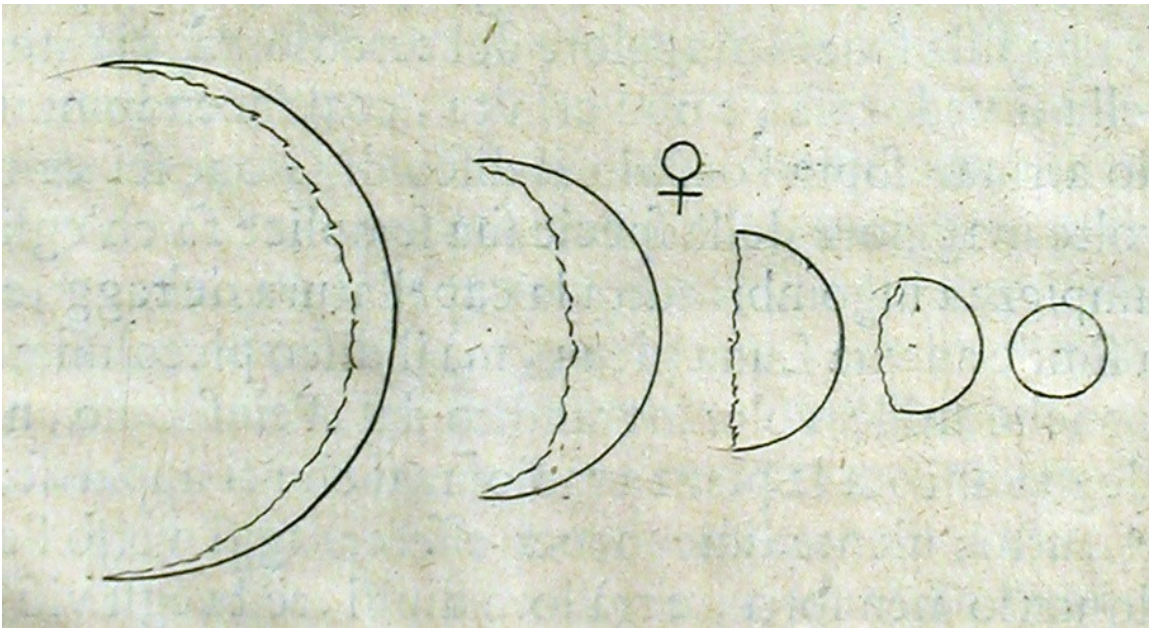

**Figure 2. Galileo's illustration of the changing phases and apparent diameter of Venus.**

Note Flamsteed's indication that the disk of Sirius is more clearly defined than that of Mercury. Flamsteed cites this observation in a discussion about the apparent diameters of fixed versus wandering stars, proceeding to argue that Mercury and Sirius were both observed with the same telescope, and the same aperture, so clearly they both had the same apparent diameter.[8]

In fact, in the case of a fixed star, the disk revealed by the telescope was false, a product of the diffraction of light through the small aperture of the telescope. On the other hand, in the case of Venus, the disk (and its phases) revealed by the telescope was true. This was surely a most difficult issue, not to be fully worked out until a satisfactory wave theory of light was developed.[9] At any rate, as Flamsteed's 1672 observation record shows, during much of this time the telescopically observed and measured disks of stars were thought to be the true bodies of those stars.

If stars seen from Earth have measurable disks, then in a Copernican universe—where the stars must be very distant in order to explain the lack of any *annual parallax* (that is, any measurable effect in the appearance of the stars as a result of Earth's orbital motion)—it necessarily follows that every visible star must be enormous. Otherwise, no visible star would show measurable size. Since apparent sizes could be measured, then every visible star must actually exceed the size of Earth's orbit. Why? Because, in the Copernican universe the size of Earth's orbit is vanishingly small compared to the starry universe. But the stars have small but measureable apparent sizes. As "small but measurable" is larger than "vanishingly small", every visible star had to be larger than Earth's orbit. This compact expression of the star size argument was put forward by J. G. Locher and Christoph Scheiner in 1614.[10]

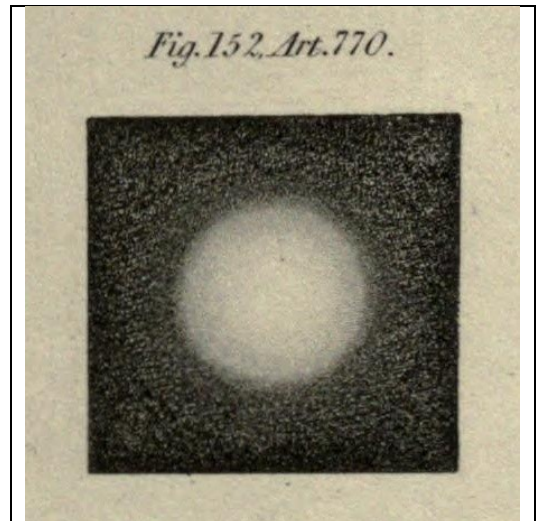

Figure 3. Illustration by John F. W. Herschel of a star as seen through a telescope of size similar to what was used for stellar observations in the seventeenth century.[7] Compare to Flamsteed's notes cited here. This disk would be considerably smaller than the disk of Venus shown in Figure 2. The diffraction of light waves through the telescope's aperture creates the globe-like appearance, greatly inflating the apparent sizes of stars. A full understanding of diffraction and the wave nature of light was not developed until the early nineteenth century.

---

[7] (Herschel 1828)
[8] (Baily 1835, 205), (Graney 2015)
[9] (Graney and Grayson 2011)

The star size argument dated as least to Tycho Brahe, well prior to Locher and Scheiner. Giovanni Battista Riccioli reinforced it with extensive telescopic observations well afterwards, as Hooke notes. And, per Hooke again, Riccioli (and Brahe) used the giant stars as an argument against the Copernican system.[11] But many Copernicans embraced the giant stars. Notable among these was Johannes Kepler, who argued that all visible stars were larger than Earth's orbit, and the most prominent ones were larger than Saturn's orbit (and thus larger than an entire geocentric universe). Kepler used the giant stars as an argument that, contra Bruno, the stars could not be other suns, orbited by other Earths—because basic observations and calculations showed that the stars were not other suns.[12]

Obviously Tacquet was using a version of the star size argument decades after Kepler, as Hooke notes. Tacquet gives the argument a slightly different flavor, however, and the flavor is a little different than what Vanpaemel describes. Tacquet notes that the baseline for parallax observations in a geocentric universe is the Earth itself—that is, the two most widely separated positions from which an astronomer can observe the stars are the opposite sides of the Earth. In a heliocentric universe, the baseline is Earth's orbit. Thus, says Tacquet, whatever the proportion was between the sizes of stars and the size of the *Earth* in the geocentric universe, that proportion must exist between the sizes of stars and the size of *Earth's orbit* in the heliocentric universe. Let us now see exactly what Tacquet has to say on this.[13]

---

[10] (Graney 2017, 30)
[11] (Graney 2015, 32-38, 129-139)
[12] (Graney 2019)
[13] (Tacquet 1668, 205-209)

# LIBER QUINTUS

## CAPVT II.

De Fixarum magnitudine, distantia, lumine, numero, speciebus, formis.

### NVMERVS XXI.

*Distantia Fixarum a centro Terrae excedit 70000, imo 100000 Semidiametrorum Terrae.*

Proinde (assumpta Terrae minima Semidiametro 3265 milliariorum Bonon.) distantia Fixarum excedit 228550000 milliariorum Bonon. hoc est 228 milliones, 550 millia: imo excedit milliar. Bon. 326500000, hoc est 326 milliones, 500 millia.

Consentiunt omnes Astronomi, vt ostendam *lib.7. num. 27.* independenter ab his, distantiam Saturni a Terra fere decuplam esse distantiae Solis a Terra. Quare cum *lib. 3. num. 19.* ostensum sit distantiam Solis numquam esse minorem 7000 Semidiametrorum Terrae, sive 22855000 milliariorum Bonon. his per 10 multiplicatis, distantia Saturni a Terra erit non minor 70000 Semid. Terr. siue 228550000 milliar. Bonon. Atqui Fixae sunt altiores Saturno, cum ab eo Eclipsentur. Ergo Fixae diftant a Terra plus quam 70000 Semid. Terr. sive 228550000. milliar. Bonon.

Deinde *num. 1*. ostendi nullam potuisse vmquam Parallaxim Fixarum obseruari: quae si esset saltem 2. secundorum, in triangulo rectangulo ABD colligeretur distantia 100000 Semidiametrorum Terrae. Igitur Fixarum distantia 100000 Semid. Terr. maior est. Neque videtur illis posse minor assignari.

# BOOK 5

## CHAPTER 2

Concerning the magnitude, distance, light, number, kinds, and forms of the Fixed Stars.

### NUMBER 21.

*The distance of the Fixed Stars from the center of Earth exceeds 70,000—no, 100,000—Semidiameters of the Earth.*

So then (from a minimum assumed Semidiameter of Earth of 3265 Bolognese miles) the distance of the Fixed Stars exceeds 228,550,000 Bolognese miles: indeed it exceeds 326,500,000 B. miles.

All of the Astronomers agree (and as I will show independently in *book 7, number 27*), the distance of Saturn from Earth to be generally ten times the distance of the Sun from Earth. Hence, since in *book 3, number 19* it is shown the distance of the Sun never to be less than 7000 Semidiameters of Earth, or [3265 × 7000 =] 22,855,000 B. miles, and this is multiplied by 10: the distance of Saturn from Earth will be not less than 70,000 Terrestrial Semidiameters or 228,550,000 B. miles. And yet the Fixed Stars are higher than Saturn, since they will be eclipsed by him. Therefore the Fixed Stars stand apart from Earth more than 70,000 Terr. Semidiam. or 228,550,000 B. miles.

Then I have shown in *number 1* that no Parallax of the Fixeds has ever been able to be observed. If that might be at least 2 seconds, in right triangle ABD (Figure 8, book 3 [below]) [where AB is a Terrestrial Semidiameter] it might compute a distance 100,000 Semidiameters of Earth. Therefore the distance of the fixeds is greater than 100,000 Semidiameters of Earth. Nor does it seem possible for them to be assigned less distance.

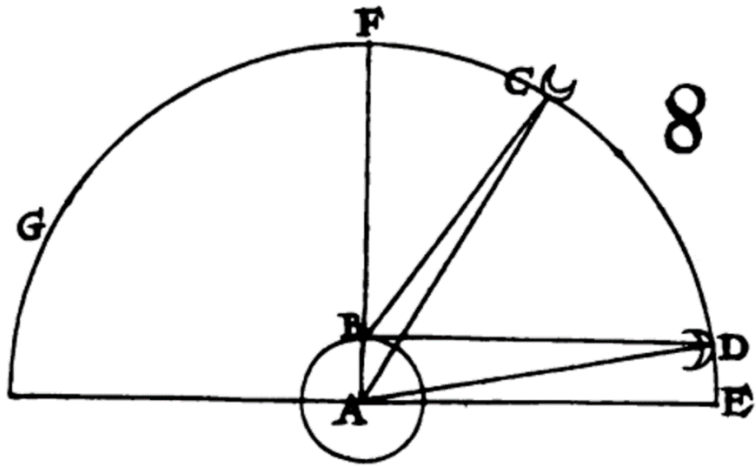

| | |
|---|---|
| Notus est igitur (quod ipsum quidem monumentum Astronomiae eximium est) distantiae terminus, infra quem Fixas non deprimi certum sit. Terminus vero distantiae alter supra quem non ascendant, explorari humano ingenio non potest. Fortassis Stellae Fixae tantum a Saturno distant, quantum ille a nobis. Credibile est etiam alias esse alijs altiores; quae & forsan est causa, cur inaequales nobis appareant. | Therefore a limit of distance is recognized (which itself is indeed an extraordinary monument of Astronomy); it is certain the Fixeds may not lie within it. The other limit of distance, beyond which they may not lie, is truly not able to be explored by human ingenuity. It may be the Fixed Stars are as much distant from Saturn, as Saturn is from us. It is even credible that some are higher than others; which perhaps is the reason why they may appear unequal to us. |

NUMBER 22 deals with the speed at which stars must circle the Earth in a geocentric universe.

NUMBER 23 discusses the apparent sizes of the stars. Tacquet includes a table (right) of apparent sizes, from Riccioli,[14] showing Sirius to measure 18″ (compare to Flamsteed's value of 15″ above), while Alcor in the handle of the Big Dipper measures just under 4½″.

| Stellæ | Qualium diam. ♄. 160. ♃. 200. | ″ | ‴ | Ordo magnitudinis. |
|---|---|---|---|---|
| Sirius | 82 | 18 | 0 | 1 |
| Lyræ lucida | 79 | 17 | 24 | 1 |
| Arcturus | 76 | 16 | 42 | 1 |
| Capella | 73 | 16 | 8 | 1 |
| Aldebaran | 70 | 15 | 24 | 1 |
| Spica | 68 | 15 | 5 | 1 |
| Regulus | 64 | 14 | 5 | 1 |
| Regel | 62 | 13 | 40 | 1 |
| Formahant | 61 | 13 | 25 | 1 |
| Antares | 60 | 13 | 12 | 1 |
| Hydra | 58 | 12 | 45 | 1 |
| Cauda ♌ | 57 | 12 | 30 | 1 |
| Procyon | 56 | 12 | 20 | 2 |
| Aquila | 50 | 11 | 0 | 2 |
| Cing. Orionis | 40 | 8 | 50 | 2 |
| Coronæ lucida | 38 | 8 | 21 | 2 |
| Polaris | 36 | 7 | 54 | 2 |
| Caput Med. | 32 | 7 | 3 | 3 |
| Propus | 28 | 6 | 10 | 4 |
| Pleias lucidior | 24 | 5 | 16 | 5 |
| Alcor | 20 | 4 | 24 | 6 |

NUMBER 24 discusses the actual physical sizes of the stars, given the apparent sizes from section number 23, and assuming the distance of 100,000 terrestrial semidiameters stated in section number 21 for a geocentric universe. Tacquet provides a table of these physical sizes (right). They range from 815 times the volume of the Earth in the case of Sirius, down to 9 times the volume of the Earth in the case of Alcor. Note again that these are the physical sizes of stars in a *geocentric* universe. Tacquet's values for individual star sizes do not agree with modern calculations, but on average

**Excedit Terram**

| | | |
|---|---|---|
| 1 | Sirius | 815 |
| 1 | Arcturus | 512 |
| 1 | Aldebaran | 402 |
| 1 | Spica | 374 |
| 1 | Regulus | 249 |
| 1 | Regel | 220 |
| 2 | Procyon | 216 |
| 2 | Aquila | 137 |
| 2 | Polaris | 62 |
| 3 | Algol | 34 |
| 4 | Propus | 26 |
| 5 | Pleias | 18 |
| 6 | Alcor | 9 |

} Vicibus

they are in error by only a few percent, as seen in the table below. Tacquet of course had to do these calculations by hand, and apparently his method introduced random errors.

---

[14] (Graney 2015, 131)

| | Star | Sirius | Arcturus | Aldeb. | Spica | Regulus | Rigel | Procyon | Aquila (Altair) | Polaris | Caput Med (Algol) | Propus | Pleias | Alcor |
|---|---|---|---|---|---|---|---|---|---|---|---|---|---|---|
| Tacquet | sec | 18 | 16.7 | 15.4 | 15.08333 | 14.08333 | 13.66667 | 12.33333 | 11 | 7.9 | 7.05 | 6.166667 | 5.266667 | 4.4 |
| Apparent | deg | 5.00E-03 | 4.64E-03 | 4.28E-03 | 4.19E-03 | 3.91E-03 | 3.80E-03 | 3.43E-03 | 3.06E-03 | 2.19E-03 | 1.96E-03 | 1.71E-03 | 1.46E-03 | 1.22E-03 |
| Diam. | rad | 8.73E-05 | 8.1E-05 | 7.47E-05 | 7.31E-05 | 6.83E-05 | 6.63E-05 | 5.98E-05 | 5.3E-05 | 3.8E-05 | 3.418E-05 | 2.99E-05 | 2.55E-05 | 2.13E-05 |
| Distance | terr. semid. | 100000 | 100000 | 100000 | 100000 | 100000 | 100000 | 100000 | 100000 | 100000 | 100000 | 100000 | 100000 | 100000 |
| **Calculated values** | | | | | | | | | | | | | | |
| Phys. Diam. | terr. semid. | 8.727 | 8.096 | 7.466 | 7.313 | 6.828 | 6.626 | 5.979 | 5.333 | 3.830 | 3.418 | 2.990 | 2.553 | 2.13318 |
| Volume | terr. vol. | 664.6 | 530.7 | 416.2 | 391.0 | 318.3 | 290.9 | 213.8 | 151.7 | 56.2 | 39.9 | 26.7 | 16.6 | 9.706946 |
| **Tacquet values** | | | | | | | | | | | | | | |
| Volume | terr. vol. | 815 | 512 | 402 | 374 | 249 | 220 | 216 | 137 | 62 | 34 | 26 | 18 | 9 |
| Phys. Diam. | terr. semid. | 9.341 | 8.000 | 7.380 | 7.205 | 6.291 | 6.037 | 6.000 | 5.155 | 3.958 | 3.240 | 2.962 | 2.621 | 2.080 |
| (calculated) | | | | | | | | | | | | | | |
| | Error vol. | 22.6% | -3.5% | -3.4% | -4.4% | -21.8% | -24.4% | 1.0% | -9.7% | 10.4% | -14.8% | -2.7% | 8.1% | -7.3% |
| | avg. | | | | | | | | | | | | | -3.8% |
| | Error diam. | 6.6% | -1.2% | -1.2% | -1.5% | -8.5% | -9.8% | 0.3% | -3.4% | 3.2% | -5.5% | -0.9% | 2.6% | -2.6% |
| | avg. | | | | | | | | | | | | | -1.7% |

NUMBER 25 deals with the volume of the geocentric universe, based on the "firmament of the stars" having a radius of 100,000 terrestrial semidiameters as stated in section number 21. While section number 21 discusses the possibility that the stars could extend back from Saturn to some depth, this calculation assumes the stars are all equidistant from Earth.

## NVMERVS XXVI.

*In Hypothesi Terrae Motae, Orbis Magnus, sive Sphaera cuius semidiameter est distantia oculi nostri vel centri Terrae a Sole, ad Firmamentum instar puncti est.*

Cum enim in Stellis Fixis numquam vlla Parallaxis fuerit deprehensa, Parallaxis autem illa siqua esset, oriri deberet ab oculi nostri distantia a centro Firmamenti; perspicuum est distantiam illam ad distantiam Fixarum esse insensibilem. Atqui in Hypothesi Terrae Motae, (hoc est, si Terra Eclipticam percurrente, Sol in Firmamenti centro quiescat) distantia oculi nostri a centro Firmamenti, est ipsa distantia oculi a Sole, hoc est Orbis Magni semidiameter. Ergo in Hypothesi Terrae Motae semidiameter Orbis Magni ad distantiam Fixarum, siue Firmamenti semidiametrum, insensibilis est. Quare cum sphaerarum proportio triplicata sit proportionis diametrorum, ipse Orbis Magnus ad Firmamentum a fortiori (vt loquuntur Philosophi) insensibilis erit.

## NUMBER 26.

*In the hypothesis of a Moved Earth, the Great Orb (or the sphere whose semidiameter is the distance from the Sun to our eye or to the center of the Earth) is, compared to the Firmament of Fixed Stars, the equivalent of a point.*

No Parallax in the Fixed Stars has ever been detected, even though such Parallax ought to arise, owing to the distance of our eye from the center of the Firmament. Therefore it is evident that distance to be insensible compared to the distance of the Fixeds. Yet in the hypothesis of the Moved Earth (that is, where the Sun may rest in the center of the Firmament, with the Earth running through the Ecliptic), the distance of our eye from the center of the Firmament is the distance of the eye from the Sun. This is the semidiameter of the Great Orb. Therefore in the Hypothesis of the Moved Earth, the semidiameter of the Great Orb is insensible compared to the distance of the Fixeds, or the semidiameter of the Firmament. Whereby the proportion of spheres may be the cube of the proportion of diameters, *a fortiori* (as the Philosophers say), the Great Orb itself will be insensible compared to the Firmament.

### Corollarium.

*Hinc sequitur Orbem Magnum esse ad Firmamentum Copernicanum, hoc est debitum Hypothesi Terrae Motae; vt Terra est ad Firmamentum commune; hoc est, debitum Hypothesi Terrae Stantis.*

Porro insensibilitas Orbis Magni ad Firmamentum alio adhuc modo demonstrabitur. Neque enim ea sola Parallaxis, quam efficit semidiameter Orbis Magni, sed etiam illa, quam, diameter tota, in Terrae Motae Hypothesi euanescit. Sit Orbis Magnus BFD [Fig. 20], Firmamentum CIG; eiusque centrum A, in quo Sol immobilis haereat, Terra motu Annuo Orbem Magnum percurrente. Eligatur Stella quaepiam cuius

### Corollary.

*Hence it follows the Great Orb is to the Copernican Firmament (that required by the Hypothesis of the Moved Earth) as the Earth is to the common Firmament (that required by the Hypothesis of the Standing Earth).*

The insensibility of the Great Orb to the Firmament will be demonstrated again, by yet another method. Not only does the Parallax which the semidiameter of the Great Orb produces vanish in the Hypothesis of the Moved Earth, but also that which the whole diameter produces. Let BFD (Figure 20 [below]) be the Great Orb; CIG be the Firmament; and A be the center of that, on which is fixed the immobile Sun, with the Altitudo Meridiana sit maior 20 Grad. ac proinde refractioni non obnoxia; & quando Terra ex. gr. est in D, obseruato Stellae appulsu ad Meridianum, vt docui *lib. 3. num. 2*, capiatur eius a Vertice distantia, angulus nempe CDI. Deinde post 6 menses, cum Terra emenso semicirculo est in B, notato rursum appulsu eiusdem Stellae ad Meridianum, inquiratur, vt prius, distantia eius a vertice CBI. Reperietur haec aequalis semper priori. Atqui reuera est maior est, cum angulus externus CBI internum CDI excedat angulo I, qui proinde vtriusque differentia est, diciturque Parallaxis Orbis Annui, causata a tota Orbis Magni diametro BD. Angulus igitur I insensibilis est: ac proinde etiam BD Orbis Magni diameter ipsum subtendens, ad Firmamenti distantiam nullam habet proportionem sensibilem.

Earth running through the Great Orb by Annual motion. Any Star may be chosen whose Meridian Altitude may be greater than 20 Degrees (and so then not liable to refraction). When Earth is at D, for example, and the approach of the Star toward the Meridian is observed, as I have taught in *book 3, number 3*, the distance of it from some Mark may be recorded, namely angle CDI. Then after 6 months, when Earth is in B, having passed through a semicircle, and the approach of the same Star toward the Meridian again is noted, the distance CBI of it from the Mark may be examined as before. This always will be found to be equal to the previous distance. And yet in truth it is larger, since the external angle CBI exceeds the internal CDI by angle I, which hence is the difference of the two, and it is called the Parallax of the Annual Orb, caused by the whole diameter of the Great Orb BD. Consequently the angle I is insensible. Hence also the diameter of the Great Orb BD subtending it holds no sensible proportion to the distance of the Firmament.

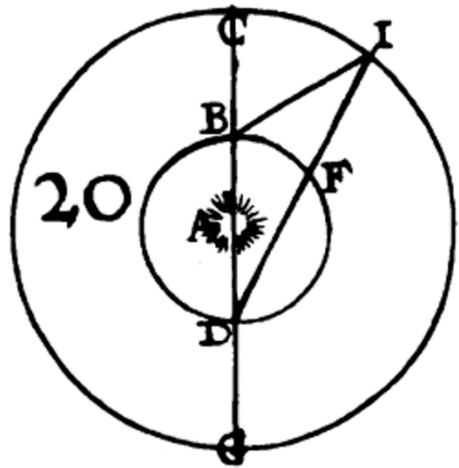

NVMERVS XXVII.

*In Hypothesi Terrae Motae, Stellae Fixae eam habent proportionem ad Orbem Magnum, quam habent in Terrae Stantis Hypothesi ad Terram.*

NUMBER 27.

*In the Hypothesis of the Moved Earth, the Fixed Stars hold the same proportion to the Great Orb, that they hold to the Earth in the Hypothesis of the Standing Earth.*

Cum Diameter Apparens cuiusuis Fixae, hoc est angulus sub quo Fixa videtur ex Terra, reperiatur praescindendo ab vtralibet Hypothesi Telluris Motae vel Stantis, vt patet ex *nu.* 23, liquet Fixam, ex. gr. *Spicam*, in vtraque Hypothesi aeque multa sui Firmamenti Minutae subtendere; tot nimirum, quot obseruatio exhibet. Igitur *Spica* Copernicaea est ad Firmamentum Copernicaeum, vt *Spica* communis est ad Firmamentum commune. Atqui Firmamentum Copernicaeum est ad Orbem Magnum, vt Firmamentum commune ad Terram. Ex aequo igitur *Spica* Copernicaea est ad Orbem Magnum, vt *Spica* communis ad Terram.

The Apparent Diameter of any Fixed (that is, the angle under which the Fixed is seen from Earth), may be found independently from either Hypothesis of the Earth (Moved or Standing). As is obvious from *number 23*, it is clear a Fixed (*Spica*, for example) to subtend in either Hypothesis equally many Minutes of its Firmament; no doubt just that many, as are recorded by observation. Therefore the Coperican *Spica* is to the Copernican Firmament, as the common geocentric *Spica* is to the common geocentric Firmament. And yet the Copernican Firmament is to the Great Orb, as the common Firmament is to the Earth. Thus the Copernican *Spica* is to the Great Orb, as the common *Spica* is to the Earth, by reason of proportion.

NVMERVS XXVIII.

*Enormis Fixarum Stellarum magnitudo in Terrae Motae Hypothesi.*

In Hypothesi Terrae Motae, Stellae Fixae sunt orbe magno maiores. Ex omnium quippe Astronomorum sententia pleraeque Fixae molem Terrae excedunt vel adaequant. Cum igitur per *num. praeced.* ita se habeant Fixae ad Orbem Magnum in Hypothesi Terrae Motae, vt eaedem sunt ad Terram in Hypothesi Terrae Stantis; manifestum est in Terrae Motae Hypothesi plerasque Fixas Orbem Magnum excedere vel aequare.

Nos autem cum Ricciolo ostendimus num. 24. *Sirium* esse maiorem Terra vicibus saltem 815; *Alcor* vero vnam e minimis, saltem nouies. Igitur in Hypothesi Terrae Motae *Sirius* 1. magn. saltem 815, *Alcor* mag. 6. saltem nouies toto Orbe Magno maior est.

Galilaeus in suo Mundi Systemate immanem istam Fixarum magnitudinem nequidquam conatur eludere a pag. 350 *vsque ad* 383 discursu longissimo, sed nihil eorum, quae *num.* 26, 27, 28, iam demonstrauimus, infirmante.

NUMBER 28.

*The enormous size of the Fixed Stars in the Hypothesis of the Moved Earth.*

In the Hypothesis of the Moved Earth, the Fixed Stars are larger than the Great Orb. Of course from the opinion of all Astronomers, most Fixeds exceed or equal the bulk of Earth. Since then through the *preceding number*, in the Hypothesis of the Moved Earth the Fixeds are to the Great Orb, as they are to Earth in the Hypothesis of the Standing Earth, it is manifest that, in the Hypothesis of the Moved Earth, most Fixeds exceed or equal the Great Orb.

But we have shown in *number 24* with Riccioli that *Sirius* is greater than Earth by at least 815 times, and *Alcor* (truly one of the small stars) by at least 9 times. Therefore, in the Hypothesis of the Moved Earth, 1st magnitude Sirius is at least 815 times, and 6th magnitude *Alcor* is at least 9 times, larger than the entire Great Orb.

Galileo in his System of the World attempts in vain to elude this monstrous magnitude of the Fixeds by a long discourse (page 350-

> 383).[15] But he weakens nothing of what we have demonstrated through *numbers 26-28*.

The discourse in Galileo's *Dialogue Concerning the Two Chief World Systems: Ptolemaic and Copernican* to which Tacquet refers is what follows Simplicio's introduction of the star size argument on the Third Day of the *Dialogue*. Simplicio introduces the argument via reference to Locher and Scheiner's discussion, mentioned above.[16] Through the character of Salviatti, Galileo gives a smaller measure for the apparent diameter of a star than does Tacquet, Riccioli, and Flamsteed, 5″ rather than roughly 15″.[17] But per Scheiner and Locher's argument about "small but measurable" being larger than "vanishingly small", the conclusion that every visible star had to be far larger than the sun still held.[18] No definite refutation of the argument is produced by Salviati, Sagredo, and Simplicio in their discussion. Scheiner and Locher's argument was simple and robust, and would only be refuted when astronomers determined that the disk of a star revealed by the telescope was false. Evidence for this would begin to accumulate in the latter part of the seventeenth century.[19]

As we have seen here, Tacquet's version of the star size argument was also simple and robust. While Hooke describes him and Riccioli as putting the argument forward "with great vehemency and insulting", Tacquet's presentation of it seen here is very straightforward. Perhaps a more vehement and insulting (and even more entertaining) version is to be found in other of his works.

---

[15] (Galilei 2001, 416-432)
[16] (Galilei 2001, 416)
[17] (Galilei 2001, 417). Galileo states in a variety of his writings over a wide span of years that stars seen through a telescope measure a few seconds of arc in diameter—see (Graney 2015, 45-49).
[18] (Galilei 2001, 430, 432)
[19] (Graney 2015, 148-155), (Graney 2019)

# WORKS CITED

Baily, Francis. 1835. *An Account of the Rev'd John Flamsteed, the First Astronomer-Royal: Compiled from His Own Manuscripts, and Other Authentic Documents.* London.

Galilei, Galileo. 2001. *Dialogue Concerning the Two Chief World Systems: Ptolemaic and Copernican.* Translated by Stillman Drake. New York: Modern Library.

Graney, Christopher M. 2017. *Mathematical Disquisitions: The Booklet of Theses Immortalized by Galileo.* Notre Dame, Indiana: University of Notre Dame Press.

—. 2015. *Setting Aside All Authority: Giovanni Battista Riccioli and the Science Against Copernicus in the Age of Galileo.* Notre Dame, Indiana: University of Notre Dame Press.

Graney, Christopher M. 2019. "The Starry Universe of Johannes Kepler." *Journal for the History of Astronomy* 50 (2): 155-173.

Graney, Christopher M., and Timothy P. Grayson. 2011. "On the telescopic disks of stars – a review and analysis of stellar observations from the early 17th through the middle 19th centuries." *Annals of Science* 68: 351-373.

Herschel, John F. W. 1828. *Treatises on Physical Astronomy, Light and Sound Contributed to the Encyclopædia Metropolitana.* London and Glasgow: Griffin and Co.

Hooke, Robert. 1674. *An Attempt to Prove the Motion of the Earth from Observations.* London.

Tacquet, André. 1668. *Opera Mathematica.*

Van Helden, Albert. 1985. *Measuring the Universe: Cosmic Dimensions from Aristarchus to Halley.* Chicago: University of Chicago Press.

Vanpaemel, G. H. W. 2003. "Jesuit Science in the Spanish Netherlands." In *Jesuit Science in the Republic of Letters*, edited by Mordecai Feingold. Cambridge, Massachusetts: MIT Press.